\documentclass{JFM-FLM_Au}
\input{myCommands}
\usepackage{bm}
\usepackage[nameinlink, noabbrev]{cleveref}
\usepackage{multicol, multirow}
\usepackage{siunitx}
\usepackage{upgreek}

\hypersetup{
colorlinks = True,
urlcolor   = blue,
citecolor  = blue,
linkcolor = black,
}

\definecolor{revcol}{RGB}{0, 0, 0}

\lefttitle{F. Kranz, D. Morón and M. Avila}
\righttitle{Minimisation of energy consumption in turbulent pipe flow}
\DeclareRobustCommand{\rev}[1]{{\color{revcol}#1\xspace}}

\title{Bayesian minimisation of energy consumption in turbulent pipe flow via unsteady driving}
\author{%
Felix~Kranz\aff{1},
Daniel~Morón\aff{1} 
\and
Marc~Avila\aff{1}$^,$\aff{2}}
\affiliation{%
\aff{1}University of Bremen, Center of Applied Space Technology and Microgravity (ZARM),\\Am Fallturm 2, 28359 Bremen, Germany.
\aff{2}University of Bremen, MAPEX Center for Materials and Processes,\\Am Biologischen Garten 2, 28359 Bremen, Germany.}
\corresau{Felix Kranz, \email{felix.kranz@zarm.uni-bremen.de}}

\begin{document}
\maketitle
\begin{abstract}

Turbulence accounts for most of the energy losses associated with the pumping of fluids in pipes. Pulsatile drivings can reduce the drag and \rev{energy consumption} required to supply a desired mass flux, when compared to steady driving. \rev{However, not all pulsation waveforms yield reductions}. Here, we compute drag- and \rev{energy}-optimal driving waveforms using direct numerical simulations and a gradient-free black-box optimisation framework. Specifically, we show that Bayesian optimisation is vastly superior to ordinary gradient-based methods \rev{in terms of computational efficiency and robustness,} due to its ability to deal with noisy objective functions, as they naturally arise from the finite-time averaging of turbulent flows. \rev{We identify optimal waveforms for three Reynolds numbers and two Womersley numbers.} At a Reynolds number of $8600$ \rev{and a Womersley number of 10}, optimal waveforms reduce \rev{total energy consumption} by 22 \% and drag by 37 \%. \rev{These reductions are rooted in the suppression of turbulence prior to the acceleration phase, the resulting delay in turbulence onset, and the radial localization of turbulent kinetic energy and production toward the pipe centre.} Our results pinpoint that the predominant, steady operation mode of pumping fluids through pipes  is far from optimal. 
\end{abstract}


\section{Introduction}\label{sec:introduction}
Turbulent pipe flows are ubiquitous in engineering, ranging from large-scale oil or gas pipelines to small-scale applications like heating pipes or fresh water supply. Pumping systems are recognized as major energy consumers globally, contributing a substantial share to the overall electrical energy demand \citep{McKeon2010}. In certain industrial plant operations, pumping systems even account for up to 50~\% of the energy usage \citep{Frenning2001}. Compared to laminar conditions, in turbulent flows, the multi-scale eddying motion is responsible for a major part of the high friction levels and ultimately the high pumping costs \citep{Blasius1913}. Hence, many efforts have been devoted to devising control strategies that can reduce drag or suppress turbulence. \rev{Successful examples go from passive strategies -- such as the use of riblets \citep{GarciaMayoral2011} -- to active strategies including the application of body forces to dampen turbulent fluctuations \citep{Marensi2019, Kuehnen2018}, as well as wall-normal blowing \citep{Mahfoze2019} and suction \citep{Mallor2023} in turbulent boundary-layer flows. A promising active strategy involves the use of oscillatory forces to reduce wall friction, such as span wise oscillations close to or at the wall \citep{Quadrio2004, Auteri2010}, or the use of unsteady pressure gradients to drive the flow. Both numerical \citep[e.g.][]{iwamoto2007, FoggiRota2023, FoggiRota2023b, Scarselli2023} and experimental \citep[e.g.][]{Souma2009, Kobayashi2021, Scarselli2023} studies have shown that certain unsteady (pulsatile) drivings can lead to significant drag reductions and energy savings. However, the space of possible flow parameters and pulsation waveforms capable of achieving such improvements is vast and remains only partially explored. 

}

\rev{Pulsatile flow} is governed by three factors. First, the (time-averaged) Reynolds number $\Rem = \Um D / \nu$, where $\Um$ is the time-averaged bulk velocity, $D$ is \rev{a characteristic length}  and $\nu$ is the kinematic viscosity. Second, \rev{the angular frequency $\omega$ of the pulsation, defined using the dimensionless Womersley number $\Womersley=(D/2)\sqrt{\omega/\nu}$} and third, the driving bulk velocity waveform \rev{(WF)} described by the Fourier expansion \par\noindent
\begin{equation}\label{eqn:fourier1}
    U(t) = \Um + \sum^\infty_{k=1} a_k \cos\left(\rev{\omega k} t\right) + \sum^\infty_{k=1} b_k \sin\left(\rev{\omega k}t\right),
\end{equation}
where $a_k, ~b_k$ are the Fourier coefficients of the pulsation.

\rev{

There are two main ways to impose a pulsatile flow: either by prescribing a time-dependent stream-wise pressure gradient or by imposing the desired bulk velocity. Most studies adopt the former approach, modulating the stream-wise pressure gradient in time. In channel flow, \cite{iwamoto2007} numerically investigated periodic square-wave pressure gradients, cyclically alternating between positive and negative values. By carefully choosing the pulsation frequency, they found that the cycle-averaged skin friction can be reduced compared to the corresponding steady flow (the one that produces the same time-averaged flow rate). Inspired by the latter, \cite{FoggiRota2023, FoggiRota2023b} employed a simple temporal waveform for the pressure gradient, consisting of a periodic on–off pumping sequence. Using direct numerical simulation (DNS), at $\Rem\approx 4600$, the best-performing waveform showed energy savings of up to $17~\%$ when accounted for the achieved flow rate. Savings were obtained when the flow spent a significant fraction of the period in a transient quasi-laminar state, where there was no axial pressure gradient and the flow rate slowly decayed.\\
\cite{Souma2009} substantiated the numerical findings of  \cite{iwamoto2007} by doing corresponding experiments in pipe flow. More recently, \cite{Kobayashi2021} did a similar experimental study at low Reynolds numbers ($\Rem \approx 3600$). They automatically generated more than 7000 different driving pressure waveforms, confirming the reduction of cycle-averaged drag. However, drag reduction is not satisfactory to achieve a reduction of net energy consumption. For example, at a mean Reynolds number of $\Rem\approx5900$ and high frequency regimes ($\Womersley\in[39,53]$), \cite{Manna2015} numerically showed that the mean turbulent friction can be reduced by harmonically modulating the driving pressure in time. However, their method required extra energy when compared to steady driving and led to a decrease of the pumping efficiency. A more complex control strategy was investigated by \cite{Ding2024} in pipe flow experiments. They found that drag can be significantly reduced over a wide range of frequencies, amplitudes, and Reynolds numbers by oscillating the pipe's surface azimuthally.
}

\rev{Inspired by the pulsatile rhythm of the mammalian heart,\cite{Scarselli2023} proposed to modulate the bulk velocity -- rather than adjusting the pressure gradient -- with a cardiac-inspired (triangular) waveform. They found that this approach could suppress key turbulent features. For instance, }in their best performing case, at $\Rem=8600$ and $\Womersley=14$, they achieved drag reductions of up to $27~\%$ and \rev{energy} savings of up to $9~\%$. The authors attributed the reductions in drag to turbulence being initially ``frozen" during acceleration phases, which minimized the impact of mean flow velocity variations on turbulent stresses. Furthermore, the reduction of \rev{wall shear stress} was linked to the delayed response of turbulence to the varying pressure gradient by \citet{Liu2024}. 
In this paper, we follow up on the work of \cite{Scarselli2023} and aim to devise periodic \rev{(bulk velocity)} waveforms that either minimize drag or \rev{energy} consumption \rev{for different Reynolds numbers $\Rem \in \{4300, 5160, 8600\}$ and for Womersley numbers of $\Womersley \in \{10, 10\sqrt{2}\}$}. In contrast to \cite{Scarselli2023} and \cite{FoggiRota2023, FoggiRota2023b}, we use optimisation algorithms in order to minimize an objective functional $\mathcal{J}$ (the turbulent drag or \rev{energy} consumption) while delivering a desired mean bulk $\Ud$, that is \par\noindent
\begin{equation}\label{eqn:opt1}
    \min_{U(t)} \ \mathcal{J}\big(U(t)\big) \quad \text{subject to} \quad \Um = \Ud.
\end{equation}
Since there is no analytical framework to understand how different waveforms influence the objective function in \cref{eqn:opt1} (i.e., $\J$ is a black-box), evaluating $\J$ requires DNS. Tackling the optimisation problem \cref{eqn:opt1} by ordinary gradient-based algorithms demands for accurate computations of $\mathcal{J}(U(t))$, generally leading to long simulations of dozens of periods. For example, \cite{Scarselli2023} obtained cycle-average values of \J by considering up to 14 periods, while \cite{FoggiRota2023, FoggiRota2023b} averaged over up to 36 periods. Therefore, to optimise \J, methods are needed that are robust to the statistical error arising from the averaging of turbulent flows.

The rest of the paper is structured as follows. In \cref{sec:methods}, we present the numerical model, the DNS setup and three finite-dimensional simplifications of \cref{eqn:opt1} that, by design, fulfil the $(\Um=\Ud)$-constraint from \cref{eqn:opt1}. In \cref{sec:methods_bo}, we describe the Bayesian optimisation method we use to find optimal waveforms \rev{(minimizers of \eqref{eqn:opt1})} and show it is vastly superior to gradient-based methods. In \cref{sec:results}, we present and discuss optimal driving waveforms and in \cref{sec:conclusion} we draw some conclusions.

\section{Methods}\label{sec:methods}
\subsection{Governing equations}
We considered the flow of a viscous Newtonian fluid with constant properties in a straight smooth rigid pipe of circular cross-section of diameter $D$ and, \rev{if not stated otherwise,} length $L=5D$. The flow is assumed to be incompressible and governed by the dimensionless Navier–Stokes equations \par\noindent
\begin{equation}\label{eqn:NSE}
    \frac{\partial \bu}{\partial t} + (\bu \cdot \nabla)\bu - \frac{1}{\Rem} \Delta \bu  + \nabla p = \fd(t) + \bff_\mathrm{b}, \quad \nabla\cdot\bu = {0},
\end{equation}
where $\bu\rev{=(u_r, u_\theta, u_z)}$ denotes the velocity field in a cylindrical coordinate system $(r, \theta, z)$, $p$ is the pressure, $\fd\rev{(t)}$ is the time-dependent axial driving force that realizes a given waveform $U(t)$ and $\bff_\mathrm{b}$ is a volumetric body force (see \cref{sec:bf}). All variables are rendered dimensionless using $D$, the time-averaged bulk velocity $\Um$ and the fluid's density $\rho_\mathrm{f}$. \rev{The dimensionless period length is given by $T=2\pi \Rem / \Womersley^2.$} We employ no-slip boundary conditions at the wall and periodic boundary conditions in the axial and azimuthal direction. \rev{For the remainder of this paper, $\overline{(\bcdot)}$ indicates temporal averages, while $\langle \bcdot \rangle_{\bm{\beta}}$ denotes spatial averaging with respect to the direction(s) $\beta.$}

\subsection{Body force \rev{to trigger turbulence transition}}\label{sec:bf}

\begin{figure}
    \centering
    \includegraphics[width=\linewidth]{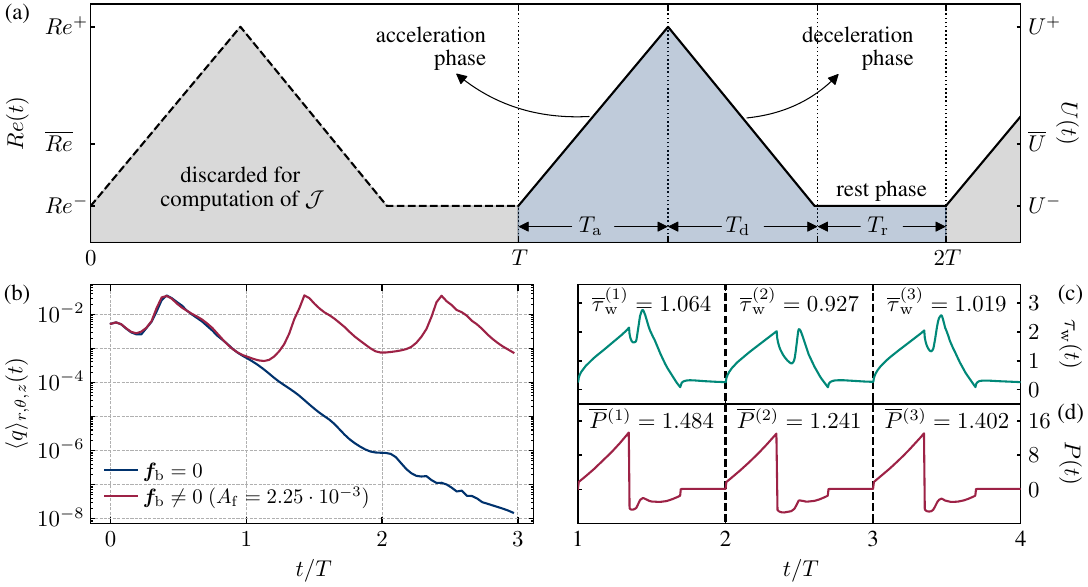}
    \caption{(a) Schematic description of the considered triangular waveforms in
terms of the time variant Reynolds number $\Re(t)$ or bulk velocity $U(t)$ (right hand side labels). (b) The evolution of the volume-integrated cross-stream turbulent kinetic energy (in units of $\Um{}^2$) over three periods of a run driven according to (a) where $\Rem=4300$, $\Re{}^+=9400$, $\Re{}^- = 1600$ and $\Ta=0.345T$. (c) The \rev{wall shear stress $\tw(t)$} and power input \rev{$P(t)$} over the last three periods of a four period run driven in the same manner as (b) \rev{where} $\Af=2.25\cdot 10^{-3}$. The \rev{wall shear stress} is normalized with respect to the steady \rev{wall shear stress} obtained by the Blasius friction and the power input accordingly \rev{(i.e. units of $\mtw{}_\mathrm{,b}$ and $\mP_\mathrm{b}$, respectively)}.}
    \label{fig:fig1}
\end{figure}

\cite{Scarselli2023} considered waveforms consisting of a constant acceleration, followed by a constant deceleration and a constant low-velocity phase. Such a waveform, used as a baseline for our calculations, is shown in  \cref{fig:fig1}(a), where \rev{$Re(t)= U(t)D/\nu$}, $Re{}^+=\max_t \Reynolds(t)=9400$, $\Reynolds{}^-=\min_t Re(t)=1600$ and $\Rem=4300$. In \cref{fig:fig1}(b), the evolution of the volume-integrated cross-stream turbulent kinetic energy \par\noindent
\begin{align}\label{eqn:q}
    \rev{\langle q\rangle_{r, \theta, z}}(t) = \frac{4}{\pi D^2 L} \int_0^L \int_0^{2\pi} \int_0^{D/2} \left(u_r^2 + u_\theta^2 \right)r \dd r \dd\theta \dd z,
\end{align}
is shown as a blue line for a direct numerical simulation initialized with a turbulent flow state (for detailed information about the simulation setup see \cref{sec:dns}). During the low-velocity phase, $\rev{\langle q\rangle_{r, \theta, z}(t)}$ decays exponentially and the flow laminarises irreversibly. \rev{This relaminarisation, differs from the experimental observations reported by \cite{Scarselli2023}, where the flow cyclically transitioned to turbulence in each period. In numerical simulations, such cyclic transition is not observed unless residual turbulence or external perturbations are present, due to the inherent linear stability of pulsatile pipe flow at these flow parameters \citep{Thomas2011}. In physical experiments, turbulence is maintained (or retriggered) by unavoidable imperfections in the setup -- such as inlet disturbances -- that are absent in idealised numerical simulations \citep{Reynolds1883, Avila2023}.\\
To model these experimental disturbances and enable a more faithful comparison with experiments, we introduced a small volumetric body force perturbation added to the right-hand side of the Navier–Stokes equations \eqref{eqn:NSE},}\par\noindent
\begin{equation}\label{eqn:forcing}
      \bff_{\mathrm{b}} = A_\mathrm{f} \left(
    \begin{array}{c}
          4r(r^2-1)^2\left[(r^2-1) \sin^2(\theta) - \left(5r^2-1\right)\cos^2\theta\right] - \left(24r^2-16\right)\sin\theta \\
 \left[-8r(r^2-1)^2\sin\theta-72r^2-16\right]\cos\theta\\
          0
    \end{array}\right), 
\end{equation}
where $A_\mathrm{f} \geq 0$ is the perturbation amplitude. Using an amplitude of $\Af=2.25\cdot 10{}^{-3}$, turbulence is triggered cyclically when the Reynolds number rises. Note that in the first period, the turbulent kinetic energy coincides with the undisturbed case, indicating that the perturbation -- while ensuring the cyclic transition to turbulence -- does not significantly alter the overall flow dynamics. \rev{This claim is later (\cref{app:robust}) verified for each of our optimal waveforms. The specific form and effect of the perturbation have been analysed in detail by \citet{kranz2024}, where different types of forcings and their amplitudes were explored.
For the remainder of this paper, we always applied the forcing \eqref{eqn:forcing} and adjusted \Af based on $\Rem$ (\cref{tab:grid}).}

\subsection{DNS}\label{sec:dns}
Simulations were carried out on a NVIDIA Tesla P100 using the open-source pseudo-spectral GPU-based research code  \href{https://github.com/Mordered/nsPipe-GPU}{\texttt{nsPipe-CUDA}} \citep{Lopez2020, Moron2024}, which uses a Fourier-Galerkin discretization ansatz for azimuthal ($N_\theta$ modes) and axial directions ($N_z$ modes) and higher-order finite differences in the radial direction ($N_r$ points). The grid is non-uniform in the radial direction with finest resolution near the walls. The value of the driving force $\fd(t)$ to enforce the bulk velocity waveform $U(t)$ is adjusted in each time step. \rev{The code integrates the Navier--Stokes equations using a Crank-Nicolson scheme and iterates on the non-linear term. The time step is adjusted dynamically so that the error between iterations on the non-linear term is always smaller than a threshold $\varepsilon$ (see \cref{tab:grid}).} \rev{Mostly, simulations were carried out at $\Womersley=10$, however, selected cases were conducted at $\Womersley=10\sqrt{2}$. Unless stated otherwise, all simulations were initialized with a turbulent initial condition obtained from a steady run.} We investigated three different average Reynolds numbers: $\Rem=4300$, $\Rem=5160$  and $\Rem=8600$. \rev{Optimizations were carried out with a relatively coarse spatial and temporal resolution, a fixed (turbulent) initial condition and a fixed body force (see \eqref{eqn:forcing}). As pointed out by \citet{FoggiRota2023}, small time steps may be needed if abrupt changes in the pressure gradient are present. In addition, they found that results are sensitive to the domain size. We verified the robustness of our optimal waveforms in terms of the spatial and temporal resolution, the initial condition, the pipe length and the forcing term and refer to the \cref{app:robust}. Detailed information about the spatial and temporal resolution is given in \cref{tab:grid}. On our fine grid, we verified a posteriori, that worst case Courant-Friedrichs-Lewy (CFL) numbers fall below 0.1 for $\Rem=5160$ and below $0.4$ for $\Rem=8600$, respectively.
} 

\begin{figure}
    \centering
    \includegraphics[width=1.0\linewidth]{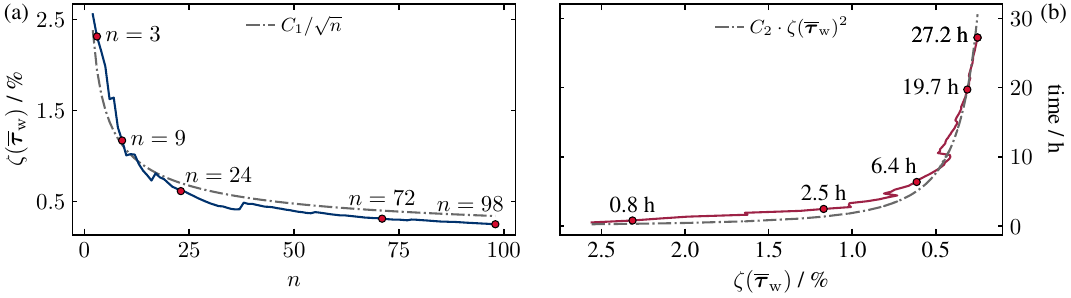}
    \caption{(a) The relative standard error of the per-period \rev{wall shear stress} ($\zeta(\overline{\bm{\tau}}_\mathrm{w})$) versus the number of averaging periods ($n$). Red dots mark the number of periods needed to achieve values for $\zeta^*$ of 2.5~\%, 1.25~\%, 0.625~\%, 0.3125~\% and 0.25~\%, respectively. The dashed line shows a $C_1/\sqrt{n}$-fit to the data. (b) The computational time (in hours) to achieve a given $\zeta(\overline{\bm{\tau}}_\mathrm{w})$, where red dots correspond to the same $\zeta^*$ values as in (a) and the dashed line shows a quadratic fit to the data.}
\label{fig:se_n}
\end{figure}

\begin{table}
    \centering
    \setlength{\tabcolsep}{3pt}
    \begin{tabular}{lcccccccccccc}
        & $\Rem$ &  $\Af$ & $N_r$ & $N_\theta$ & $N_z$ & $\Re_\tau$ & $\Delta r^+_-$ & $\Delta r^+_+$ &  $\Delta (R\theta)^+$ & $\Delta z^+$ & $\Delta t_-$ & $\Delta t_+$\\  \hline
        \multirow{3}{*}{\shortstack{optimisation \\ iterations}} & 4300 & $2.25\cdot10^{-3}$ & 80 & 192 & 270 &  277 &0.09 & 5.1 & 9.1 & 10.2 & $1.3\cdot10^{-3}$ & $1.6\cdot10^{-2}$ \\
        & 5160 & $1.15\cdot10^{-3}$  & 96 & 192 & 300 & 258 & 0.05 & 4.0 & 8.5 & 8.6 & $1.2\cdot10^{-3}$ & $1.4\cdot10^{-2}$ \\
        & 8600 & $0.5\cdot10^{-3}$  & 128 & 246 & 420 & 431 & 0.06 & 5.0 & 11.0 & 11.0 & $0.7\cdot10^{-3}$ & $1.5\cdot10^{-2}$ \\ \hline
        \hline
        WF 1 (coarse) & 5160 & $1.15\cdot10^{-3}$  & 96 & 192 &  300 & 215 & 0.05 & 3.3 & 7.0 & 7.1 & $1.3\cdot10^{-3}$ & $1.3\cdot10^{-2}$ \\
        WF 1 (fine)& 5160 & $1.15\cdot10^{-3}$  & 120 & 436 &  390 & 216 & 0.03 & 2.6 & 3.1 & 5.5 & $1.6\cdot 10^{-4}$ & $1.4\cdot 10^{-3}$  \\
        \hline
        WF 2 (coarse) & 5160 & $1.15\cdot10^{-3}$ & 96 & 192 &  300 & 222 & 0.05 & 3.4 & 7.3 & 7.4 & $1.9\cdot10^{-3}$ & $1.2\cdot10^{-2}$\\
        WF 2 (fine)& 5160 & $1.15\cdot10^{-3}$  & 120 & 436 &  390 & 222 & 0.03 & 2.7 & 3.2 & 5.7 & $2.4\cdot10^{-4}$ & $1.3\cdot10^{-3}$ \\
        \hline
        WF 3 (coarse) & 5160 & $1.15\cdot10^{-3}$  & 96 & 192 &  300 & 225 & 0.05 & 3.4 & 7.4 & 7.5 & $1.3\cdot10^{-3}$ & $1.5\cdot10^{-2}$  \\
        WF 3 (fine)& 5160 & $1.15\cdot10^{-3}$  & 120 & 436 &  390 & 232 & 0.03 & 2.9 & 3.3 & 5.9 & $1.2\cdot 10^{-4}$ & $2.8\cdot 10^{-3}$  \\
        \hline
        WF 4 (coarse) & 8600 & $0.5\cdot10^{-3}$  & 128 & 246 &  420 & 375 & 0.05 & 4.3 & 9.4 & 9.5 & $7.3\cdot10^{-4}$ & $1.3\cdot10^{-2}$  \\
        WF 4 (fine)& 8660 & $0.5\cdot10^{-3}$  & 120 & 436 &  390 & 366 & 0.03 & 3.4 & 3.7 & 5.8 & $2.9\cdot10^{-4}$ & $6.5\cdot10^{-3}$  \\
         \hline
        WF 5 (coarse) & 5160 & $1.15\cdot10^{-3}$  & 96 & 192 &  300 & 247 & 0.06 & 3.7 & 8.1 & 8.2 & $1.5\cdot10^{-3}$ & $0.8\cdot10^{-2}$  \\
        WF 5 (fine)& 5160 & $1.15\cdot10^{-3}$  & 120 & 436 &  390 & 247 & 0.04 & 3.0 & 3.6 & 6.3 & $2.4\cdot 10^{-4}$ & $1.0\cdot 10^{-3}$  \\
    \end{tabular}
    \caption{ \rev{In columns, from left to right: time averaged Reynolds number $\Rem$, forcing amplitude $\Af$, number of physical grid points in radial, azimuthal and axial direction ($N_r$, $N_\theta$, $N_z$), maximum friction Reynolds number ($\Re_\tau=\max_t\sqrt{\tau_\mathrm{w}(t)/\rho_\mathrm{f}} (D/\nu)$), minimum/maximum radial, azimuthal and axial resolution in inner units ($\Delta r^+_-$, $\Delta r^+_+$, $\Delta(R\theta)^+$ and $\Delta z^+$) and minimum and maximum time step ($\Delta t_-$, $\Delta t_+$) in units of $D/\overline{U}$. Optimisations are carried out on a coarse mesh with a large time step ($\varepsilon\approx10^{-9}$) where in sub-table ``optimisation iterations", we report the averages over all iterations. The following sub-tables display the resolutions in the optimal waveforms (WF 1--5) as computed in the optimization loop (coarse) where $\varepsilon\approx10^{-9}$ and verification cases for the optimal waveforms (fine) on a fine grid with a smaller time step ($\varepsilon\approx10^{-13}$).}}
    \label{tab:grid}
\end{table}
\rev{
We considered two quantities of interest: the $n$-cycle-averaged wall shear stress $\mtw$ and power input by $\mP$, defined as \par\noindent
\begin{equation}\label{eqn:functionals}
    \begin{aligned}
    \mtw=\frac{1}{n} \sum_{i=1}^n \mtw^{(i)}, \quad &\mtw^{(i)} = \frac{1}{T} \int_{iT}^{(i+1)T}\tw(t)\dt \\
    \mP=\frac{1}{n} \sum_{i=1}^n \mP^{(i)}, \quad &\mP^{(i)} = \frac{1}{T}\int_{iT}^{(i+1)T} Q(t)\, \Delta p(t)\dt
    \end{aligned}
\end{equation}

\noindent where $\tw(t):= \mu\langle (\partial u_z/\partial r)|_{r=D/2}\rangle_{\theta, z}(t)$ is the axial and azimuthal averaged wall shear stress, $\mu$ the dynamic viscosity, $Q(t)=(\pi/4)U(t)$ the volumetric flow rate and $\Delta p(t)$ the pressure drop. We refer to 
$\mtw{}^{(i)}$ and $\mP{}^{(i)}$ as the per-period wall shear stress and power input of period $i$, respectively. Note that the zeroth cycle ($i=0$ in \eqref{eqn:functionals}) always features significantly higher turbulence levels due to the initial condition and is always discarded for the analysis. We remark that minimizing the average power input is equivalent to minimizing the total energy consumption $nT\cdot \mP$. Thus, in what follows, power savings can be interpreted as energy savings and vice-versa.} The quantities $\mtw$ and $\mP$ were normalized with respect to the steady state values \rev{(indicated by $(\bcdot)_\mathrm{b}$)}, inferred from the Blasius friction law $2\overline{\tau}_\mathrm{w,b} = 0.0791{\Um}\  {\Rem{}^{-1/4}}$ \citep{Blasius1913}. \rev{Following \cite{Scarselli2023}}, we defined the drag reduction $D_\mathrm{r}$ and power saving $P_\mathrm{s}$ as \par\noindent
\begin{equation}\label{eqn:DrPs}
    D_\mathrm{r} \rev{= \frac{\mtw{}_\mathrm{,b}-\mtw}{\mtw{}_\mathrm{,b}}}=1-\mtw, \quad P_\mathrm{s}\rev{ = \frac{\mP_\mathrm{b}-\mP}{\mP_\mathrm{b}} }=1-\mP\rev{,}
\end{equation}
\rev{where values larger/smaller than zero indicate reduced/increased drag or power consumption, when compared to steady driving.} Because of  the chaotic nature of turbulence, a single period of a pulsatile run is not representative for the overall behaviour: \cref{fig:fig1}(c) shows the evolution of the \rev{wall shear stress} and the power input over three periods for a turbulent flow periodically driven according to the waveform in \cref{fig:fig1}(a). Analysing the per-period \rev{wall shear stress for this specific run unveils} that $\mtw{}^{(1)}$ is roughly 14.7~\% larger than  $\mtw{}^{(2)}$ and 4.4~\% larger than $\mtw{}^{(3)}$. For the power input, $\mP{}{}^{{(1)}}$ is approximately 19.5~\% larger than $\mP{}^{{(2)}}$ and 5.8~\% larger than $\mP{}^{{(3)}}$ (\cref{fig:fig1}(d)). This motivates the question of finding the minimum number of averaging periods that approaches the statistically steady values of $\mtw$ and $\mP$. \rev{To address this question, } we considered the relative standard error of the vector of per-period \rev{wall shear stress} $\overline{\bm{\tau}}_\mathrm{w}=(\mtw{}^{{(i)}})_{i=1,\ldots, n}$ and power input $\overline{\bm{P}}=(\mP{}^{(i)})_{i=1,\ldots, n}$, that are \par\noindent
\begin{equation}\label{eqn:ste}
    \zeta(\overline{\bm{\tau}}_\mathrm{w}) = \frac{\sigma(\overline{\bm{\tau}}_\mathrm{w})}{\mtw\sqrt{n}}, \quad \zeta(\overline{\bm{P}}) = \frac{\sigma(\overline{\bm{P}})}{\mP\sqrt{n}},
\end{equation}
where $\sigma(\bcdot)$ denotes the sample standard deviation \rev{and is required to fall under a threshold $\zeta^*$.} Initially, simulations were conducted and post-processed averaging over \rev{three} periods, and were automatically extended by another period until $\zeta(\bcdot) \leq \zeta^*$. Reducing $\zeta^*$ comes with extensive computational efforts: based on the simulation disseminated in \cref{fig:fig1}(c). In \cref{fig:se_n}(a) we illustrate the number of averaging periods needed to achieve a given $\zeta^*$ (computed on $\overline{\bm{\tau}}_\mathrm{w}$). Realizing $\zeta^*=2.5~\%$ requires three averaging periods, corresponding to a computational time of roughly 48 minutes (see \cref{fig:se_n}(b)). Halving $\zeta^*$ requires nine periods (2.5 hours of computing time) while reducing it by a factor of ten requires for almost 100 periods (27.2 hours).

\subsection{Finite-dimensional optimisation problem and waveform design}\label{sec:waveforms}

In the optimisation problem, \cref{eqn:opt1}, we sought bulk velocity waveforms $U(t)$ that deliver a desired averaged velocity bulk $\Ud$, while minimizing \rev{either the mean wall shear stress ($\J = \mtw$) or the mean power input ($\J=\mP$)} (see \cref{eqn:functionals}). \Cref{eqn:opt1} is a complex optimisation problem: it is constrained by a partial differential equation ($\bu$ has to fulfil the \rev{Navier--Stokes equations} \eqref{eqn:NSE}), it features state constraints (the desired flux $\Ud=1$ needs to be satisfied) and it is of infinite dimensions (seeking functions $U\rev{(t)}$). To simplify the optimisation problem \eqref{eqn:opt1} and overcome the last two complexity features, we considered two different waveforms \rev{$U_\bv(t)$} that can be defined by a finite number of parameters $\bv\in\mathbb{R}{}^{d}$, and (by design) deliver $\Ud=1$.

First, we considered triangular waveforms as in \cite{Scarselli2023}. The waveforms consists of an acceleration phase of length $\Ta$, accelerating the flow from a minimum Reynolds number ($\Re^-$) to a maximum Reynolds number ($\Re^+$), a deceleration phase of length $\Td$ slowing down it to $\Re^-$ and a rest phase of length $\Tr$ of steady driving, see \cref{fig:fig1}(a). We parametrized the waveform by the acceleration time and maximum and minimum Reynolds number ($\bv=(\Ta, \Re^+, \Re^-)$), obtaining the shape \par\noindent
\begin{align}\label{eqn:waveform}
U_{(\Ta, \Re^+, \Re^-)}(t) = \frac{\Um}{\Rem} \cdot  \begin{cases}
    \Re^-+\frac{\Re^+-\Re^-}{\Ta} t \quad &0\leq t < \Ta, \\
    \Re^+ + \frac{\Re^- - \Re^+}{\Td} (t-\Ta) \quad &\Ta\leq t < \Ta + \Td,\\
    \Re^- \quad &\Ta + \Td \leq t \leq T.\\
\end{cases}
\end{align}
By restricting the waveform in terms of fixing $\Re^+$ and $\Re^-$, this triangular waveform can be simplified into a uni-variant one, only dependent on the acceleration time $\bv=\Ta$. Note that $\Td$ is adjusted based on $\Ta = \Ta(\Td)$, so that the desired mass flux is satisfied, while $\Tr$ is fixed by $\Reynolds^+$ and $\Reynolds^-$.\\
Second, we used a truncated Fourier series to express the waveform, allowing for a wide range of continuous waveform shapes. Incorporating the desired bulk velocity $\Ud=1$ where we fix the phase of the first mode ($b_1=0$) to avoid waveforms that are identical up to a time shift. The waveform is given by \par\noindent
\begin{align}\label{eqn:fourier}
    U_\bv(t) = 1 + \sum^N_{k=1} a_k \cos\left(\frac{2\pi}{T}nt\right) + \sum^N_{k=2} b_k \sin\left(\frac{2\pi}{T}nt\right),~\bv := (a_1, \ldots, a_N, b_2, \ldots b_N).
\end{align}
In other words, we aimed to identify the $2N-1$ Fourier coefficients, such that the bulk velocity waveform minimises the \rev{mean wall shear stress} or \rev{mean} power input. 
 
Both waveform designs automatically satisfy the desired-flux-constraint in \cref{eqn:opt1} and the optimisation problem can be reduced to finite dimensions as \par\noindent
\begin{equation}\label{eqn:opt3}
    \min_{\bv \in \mathcal{Q} \subset\mathbb{R}^d } \ \mathcal{J}\big(U_\bv(t)\big),
\end{equation}
where, $\mathcal{Q}\subset\mathbb{R}^d$ is the admissible set realizing bounds for the parameter vector, specified later.  

\section{Optimisation method}\label{sec:methods_bo}

A challenge in solving the optimisation problem~\cref{eqn:opt3} is that the functional $\mathcal{J}$ is noisy. The noise amplitude is inversely proportional to the square root of the number of periods considered in the DNS, see \cref{fig:se_n}(a) \rev{and \cref{eqn:ste}}. As shown in \cref{app:gds}, gradient-based methods are not well suited for this optimisation problem owing to the large number of periods needed to reduce the noise level, so the robust evaluation of gradients of $\mathcal{J}$ is ensured. Bayesian optimisation (BO) naturally embraces the noisy nature of~\cref{eqn:opt3} and, as shown below, produces results that are consistent regardless of the number of periods chosen to evaluate  $\mathcal{J}$. 

BO  is a global data-driven optimisation technique for black-box functions that feature stochastic noise, are costly to evaluate and possibly non-convex and non-differentiable \citep{Kushner1964, Mockus1972}. After initializing a dataset by observing the objective at initial points, $\mathcal{D}{}^{(0)}=\bigcup_i\{(\bv{}^{(0)_i}, \mathcal{J}(\bv{}^{(0)_i}))\}$, where $\bv{}^{(0)_i}$ denote the initial points, a probabilistic surrogate model, $\Js{}^{(0)}$, usually in the form of Gaussian processes (GP), is fitted to $\mathcal{D}{}^{(0)}$. An acquisition function $\alpha(\bv)$ is then used to decide where to evaluate the function next by balancing the trade-off between exploration (sampling where the uncertainty of the GP is high) and exploitation (sampling where the surrogate model predicts low values). A common choice for $\alpha$ (chosen in this study) is the expected improvement, $\alpha(\bv)=\mathbb{E}[\max(0, \mathcal{J}^* - \mathcal{J}(\bv))]$, where $\mathcal{J}^*$ is the current best observed value. The next point to sample $\mathcal{J}$ is given by $\bv^{(j+1)}=\arg\max_{\bv \in \mathcal{Q}} \alpha(\bv)$ and we set $\mathcal{D}{}^{(j+1)}=\mathcal{D}{}^{(j)}\bigcup (\bv^{(j+1)}, \mathcal{J}(\bv^{(j+1)}))$. Iteratively, an updated surrogate $\Js{}^{(j+1)}$ is fitted to the updated dataset, a new point maximizing $\alpha$ is sampled and the dataset is extended. Detailed descriptions of BO  are given in \cite{Mockus1994} and \cite{Jones1998} and references therein.

For the BO algorithm, we constructed the initial dataset $\mathcal{D}^{(0)}$ by randomly sampling $5d$ initial data points, in other words, $\mathcal{D}^{(0)} = \bigcup_{i=1}^{5d} \big\{\left(\bv{}^{(0)_i}, \mathcal{J}(\bv{}^{(0)_i})\right) \big \}$, where $\bv{}^{(0)_i}$ is drawn from a continuous uniform distribution. Convergence criteria in BO are not as straightforward as in gradient-based methods, as, due to the exploring-exploiting approach, the iterative function values are generally not monotonically converging towards the optimum. As the dataset grows, the surrogate \Js is expected to improve continuously, and so the expected minimum (the minimum of \Js) is gradually refined. However, if the acquisition function favours exploring new areas, and the observed value falls out of line with the existing surrogate, expected minima may change. In order to capture that the algorithm converges to an exploration-unbiased minimum, we demanded the relative difference between expected minima to fall beneath a threshold $\epsilon$, \par\noindent
\begin{equation}\label{eqn:BO_conv}
    \left| \frac{\min_\bv \Js^{(j+1)}(\bv) - \min_\bv \Js^{(j)}(\bv)}{\min_\bv \Js^{(j)}(\bv)} \right | \leq \epsilon
\end{equation}
for a patience of at least 5 iterations. Note that this way, an optimisation is at least $5(d + 1)$ iterations long, however the risk of aborting prematurely is reduced. Convergence criteria of this kind have been reported in the literature and have been demonstrated to yield reasonable timings to stop the BO algorithm \citep{ishsibashi2023}.

We assessed the performance in terms of feasibility, computational effort and robustness of BO using the restricted (uni-variant) triangular waveform described in \cref{sec:waveforms}.  Here, choosing \rev{$\Womersley=10$,} $\Rem=4300$, $\Re^+=9400$ and $\Re^-=1600$ (see \cref{fig:fig1}(a)), we aimed to identify the $\mtw$-optimal acceleration time. The acceleration time is bounded by $0.01T$ and $0.68T$ ($\Ta \in \mathcal{Q}=[0.01, 0.68]T$) in order to obtain continuous waveforms and to ensure that $\Rem$ is realized.

\begin{figure}
    \centering
    \includegraphics[width=\linewidth]{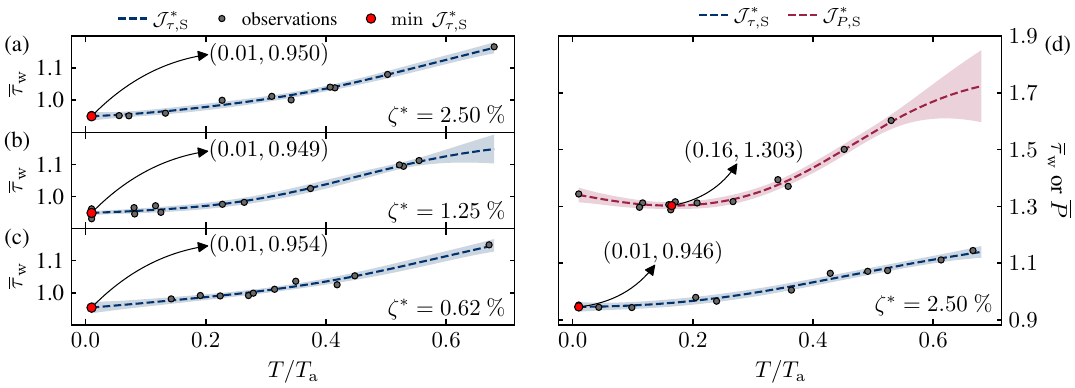}
    \caption{Panes {(a)}--{(c)} show the best surrogate for the mean \rev{wall shear stress} ($\rev{\Jst}^{*}$) and expected optima for different allowable standard errors $\zeta^*\in\{2.5, 1.25, 0.625\}~\%$. Pane {(d)} shows best surrogates for the \rev{wall shear stress} and power input, $\rev{\Jst}^*$ and $\rev{\Jsp}^*$, respectively, as well as expected optima, when reducing the number of initial averaging periods to two. In all cases, the flow was driven according to the waveform from \cref{fig:fig1}(a), where $\Rem=4300$, $\Re^-=1600$ and $Re^+=9400$. \rev{Shaded areas indicate the uncertainty of the surrogate model. Wall shear stresses are given in units of $\mtw{}_\mathrm{,b}$ and power inputs in $\mP_\mathrm{b}$.}}
\label{fig:bo1}
\end{figure}

In \cref{fig:bo1}{(a)}--{(c)}, we show the best surrogates (surrogates based on the last dataset) for the \rev{wall shear stress ($\Jst^{*}$)} for three different values of $\zeta^*$. In the Bayesian optimisation, regardless of the admissible standard error, expected optima (red markers) are obtained at the lower bound of $\mathcal{Q}$ ($\Ta=0.01T$) and surrogate models visually coincide. Using a 95~\% of confidence interval (CI), the differences in the expected function values are statistically insignificant (0.950 and 0.949 are within the 95~\% CI of 0.954 with a standard error of $\zeta=0.625~\%$). We tested the performance of BO, by further reducing the number of \rev{initial} averaging periods to \rev{$n=2$}. We refrained from decreasing it to one period only as the $\zeta^*$-criterion does not apply in that case and thus, no upper bound for the noise can be given. As shown in \cref{fig:bo1}{(d)}, the surrogate for $\mtw$ as well as the expected minimum of $\mtw=0.946$, are almost identical to cases with a higher number of averaging periods \cref{fig:bo1} {(a)}--{(c)}. In summary, unlike gradient-based methods (see \cref{app:gds}, \cref{fig:slsqp}), BO produces results that are consistent as $\zeta^*$ is reduced. 

\rev{In \cref{fig:bo1}(d), we also show the best surrogate for the mean power input ($\Jsp^*$)}. Here, the surrogate close to the \rev{expected} minimum \rev{of $\mP\approx1.3$} at $\Ta=0.16$ is flat, which intensifies the problem of accurately approximating gradients.

To put computational efforts into perspective: the optimisations shown in \cref{fig:bo1}{(d)} took about 9 hours, whereas to arrive at a similar minimum with the gradient-based SLSQP method described in \cref{app:gds} took 4.5 days. 

These results show that BO is a powerful \rev{tool} to approach the optimisation problem at hand: (i) it can deal by design with the noisy setting; (ii) it is efficient in terms of the number of function evaluations; (iii) already made observations are reusable. Hence, for all optimisations in this paper, BO was the \rev{tool} of choice. Specifically, we used the implementation of BO available in the python library \href{https://scikit-optimise.github.io/stable/}{\texttt{scikit-optimize}}.

\section{Results}\label{sec:results}
\subsection{Uni-variant approach}\label{sec:res_uni}

\begin{figure}
    \centering
    \includegraphics[width=\linewidth]{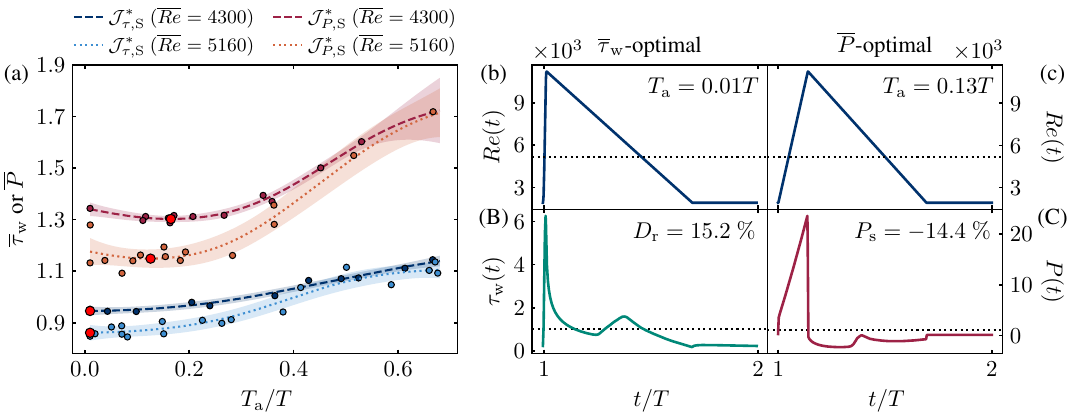}
    \caption{{(a)} Shows the best surrogates and expected minima for the mean \rev{wall shear stress} $\rev{\Jst}^*$ and the mean power input $\rev{\Jsp}^*$ at Reynolds numbers of $\Rem=4300$ and $\Rem=5160$, respectively. \rev{Shaded areas indicate the uncertainty of the surrogate model.} The plots {(b)}--{(C)} show the $\mtw$- and $\mP$-optimal and sub-optimal waveforms ($\Rem=5160$) and the resulting evolutions of the \rev{wall shear stress} and power over the time span of two periods, respectively (where capital letters are associated with the lower case letters). The average Reynolds number as well as steady values for the \rev{wall shear stress} and power input, obtained by Blasius' friction law, are indicated by dotted lines. \rev{Wall shear stresses are given in units of $\mtw{}_\mathrm{,b}$ and power inputs in $\mP_\mathrm{b}$.}}
\label{fig:bo2}
\end{figure}

\rev{At $\Womersley=10$}, we optimized the uni-variant waveform at Reynolds numbers, $\Rem=5160,~\Re^+=11280,~\Re^-=1920$, averaging initially over two periods and enforcing $\zeta^*=2.5~\%$. In \cref{fig:bo2}{(a)}, we compare the \rev{(best)} surrogates for the mean \rev{wall shear stress} \rev{($\rev{\Jst}^*$)} and the mean power input \rev{($\rev{\Jsp}^*$)} at $\Rem=4300$ and $\Rem=5160$. Again, minima of $\mtw$ are obtained at the smallest admissible acceleration time, $\Ta=0.01T$, whereas the  drag reduction improves from $\Dr=5~\%$ at $\Rem=4300$ to $\Dr=14~\%$ at $\Rem=5160$. $\mP$-optima deviate slightly in terms of the acceleration time ($\Ta=0.16T$ at $\Rem=4300$ and $\Ta=0.13T$ at $\Rem=5160$), whereas the power loss is decreased from $\Ps=-30~\%$ to $\Ps=-14~\%$. Both surrogates for $\Rem=5160$ show increased noise levels compared to the lower $\Rem$ case and thus the uncertainty of each surrogate (shaded areas) also increases. The average standard deviation of single observations regarding the surrogate increases from $0.7~\%$ to $1.3~\%$ for the $\mtw$ surrogate and from $0.8~\%$ to $2.1~\%$ for the $\mP$ surrogate. Considering the noise, the difference in expected optimal points for $\mP$ is insignificant: evaluation of the $\mP$-surrogate for $\Rem=5160$ at $\Ta=0.16$ yields a power loss of $15.0~\%$ (versus $14.4~\%$ at $\Ta=0.13$). The increased noise levels underscore that gradient-based methods are inappropriate for the given problem.

In \cref{fig:bo2}{(b)}, we show the $\mtw$-optimal driving waveform ($\Ta=0.01T$), as well as the evolution of the resulting \rev{wall shear stress} and power input, averaged over five periods. At $\zeta=2.1~\%$, the drag reduction that is actually achieved is insignificantly larger than the one projected by the surrogate ($\Dr=15~\%$ compared to $\Dr=14~\%$). Due to abrupt acceleration, peak values for the \rev{wall shear stress} overshoot steady values by up to seven times in the beginning of the period. During deceleration, the \rev{wall shear stress} quickly declines, eventually rising again until hitting a second peak at roughly $0.5T$. This behaviour may be explained as follows. In low velocity phases ($\Rem<5160$), turbulence levels decay, setting a favourable initial condition for the following acceleration. Opposed to slow accelerations, where turbulence has sufficient time to rise until being fully developed at peak velocities and realizing accordingly large \rev{wall shear stress}es, in fast accelerations, peak turbulence levels are attained during deceleration, where flow velocities are already comparably small.

A reduction of the mean \rev{wall shear stress} does not ensure that the \rev{mean} power input is lower compared to steady conditions. In fact, the negative power saving of $\Ps=-14~\%$ achieved by the a-posteriori simulation of the $\mP$-optimal waveform ($\Ta=0.13T$) in \cref{fig:bo2}(b) shows that, at this Reynolds \rev{and Womersley} number, any pulsatile driving of this shape results in power losses. The power loss compared to the steady case is caused by the additional energy input required to accelerate the flow. Accelerating the flow within short time frames requires a large pressure gradient and hence the required \rev{instantaneous} power input increases \rev{up to 24 times the steady value}.

\subsection{Tri-variant approach}

\begin{figure}
    \centering
    \includegraphics[width=\linewidth]{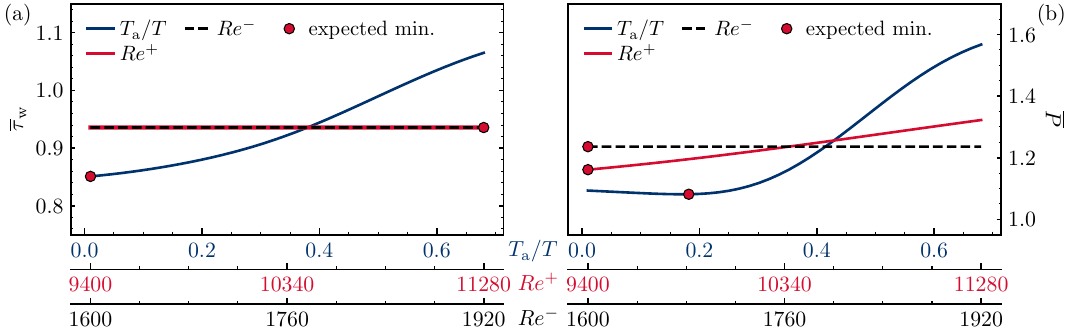}
    \caption{Partial dependence plots for the \rev{wall shear stress in units of $\mtw{}_\mathrm{b}$} (a) and the power input \rev{in units of $\mP_\mathrm{b}$} (b) in the tri-variant optimization. Circles show the expected minimum.}
\label{fig:botri}
\end{figure}

We kept $(\Rem, \Womersley)=(5160,10)$ and now allowed maximum and minimum Reynolds numbers $\Re^+\in [9400, 11280]$ and $\Re^-\in [1600, 1920]$ to change (tri-variant approach) \rev{-- a combination of the uni-variant case at $\Rem=4300$ and $\Rem=5160$}. In \cref{fig:botri}, the optimization outcome is displayed using partial dependence plots. The partial dependence is calculated by averaging the objective value for a number of random samples in the search-space, while keeping the remaining dimensions fixed \citep{goldstein2014}. This averages out the effect of varying the other dimensions and shows the influence of one or two dimensions on the objective function.

\Cref{fig:botri}(a) shows the dependence of the mean \rev{wall shear stress} on $\Ta$, $\Re^+$ and $\Re^-$. The function value is mostly ruled by the acceleration time as nearly no dependence on the maximum and minimum Reynolds number can be observed (equal values up to the fifth decimal place at lower and upper bounds for $\Re^+$ and $\Re^-$, respectively). Thus, same drag reductions of roughly $15~\%$ as in the uni-variant case are obtained at the upper bounds of $\Re^+$ and $\Re^-$ and the lower bound of $\Ta$. As shown in \cref{fig:botri}(b), the mean power input shows dependencies on the maximum Reynolds number and the acceleration time. The $\Ta$-dependence is more substantial, spanning power losses from $-8~\%$ to $-58~\%$. Varying $\Re^+$ also has an impact on $\mP$. Specifically, modifying $\Re^+$ for a fixed acceleration time can influence the power saving by up to $16~\%$ ($\mP=1.14$ for $Re^+=9400$ and $\mP=1.30$ for $Re^+=11280$ (red curve in \cref{fig:botri}\textbf{(b)}). The largest power saving of ($\Ps=-2~\%$, power loss of 2~\%) is found at an acceleration time of $\Ta=0.18T$ and at lower bounds for maximum and minimum Reynolds numbers ($\Re^+=9400$, $\Re^-=1600$), corresponding to the shortest possible rest phase. As argued before, the slight difference in optimal acceleration times ($\Ta=0.16T$ for the uni-variant and $\Ta=0.18T$ for the tri-variant case) is insignificant considering the flatness of the $\Ta$-dependence around the minimum in combination with the noisy setting. The expected minimum of $\mP=1.02$ falls in line with the a-posteriori evaluation at the optimal parameters ($\mP=1.015$, not shown here).
 
\subsection{Truncated Fourier-series}

In the Fourier-series expansion of arbitrary waveforms, we only optimized for the power input \rev{($\J=\mP$)} for computational reasons. However, we note that all found power-reducing waveforms \rev{also} reduce \rev{the mean wall shear stress} substantially. \rev{We first considered $\Womersley=10$ and later extended the analysis to  $\Womersley=10\sqrt{2}$}. At $\Rem=5160$, we considered five coefficients $a_1,a_2,a_3$ and $b_2, b_3$ in \cref{eqn:fourier} and in order to restrict the maximum Reynolds number, we limited the magnitude of each by $1/6$ ($|a_k|, |b_k|\leq 1/6 $, $k=1,\ldots, N$), because in the BO toolkit used, it is not trivial to realize non-box constraints (i.e. to restrict \rev{$\Re{}^+$} directly). 

\begin{figure}[h!]
    \centering
    \includegraphics[width=\linewidth]{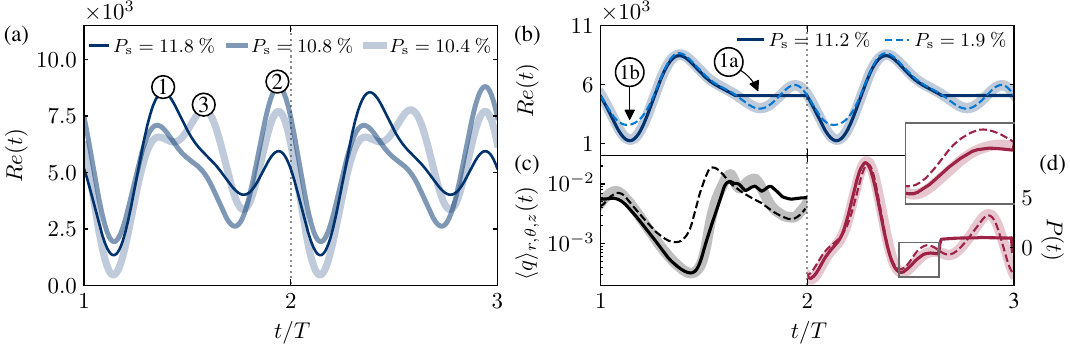}
    \caption{(a) Power-optimal waveforms obtained from three independent runs of the truncated Fourier approach at $\Rem=5160$ \rev{and $\Womersley=10$} where $N=3$ (see \eqref{eqn:fourier}) and $|a_k|, |b_k|\leq 1/6$, $k=1, 2, 3$ (phase-adjusted for the minium velocity). (b) best-performing waveform from (a) (\rev{WF} 1) and two modifications thereof (1a and 1b). (c) and (d) evolution of the cross-sectional kinetic energy \rev{(units of $\Um{}^2$)} and of the power input (\rev{units of $\mP_\mathrm{b}$;} line styles according to (b)).}
\label{fig:f_opt}
\end{figure}

\Cref{fig:f_opt}(a) shows one period of the $\mP$-optimal waveforms for three independent optimization runs (labelled 1--3). On average, four periods were necessary to achieve $\zeta^* = 2.5~\%$ and 46 iterations of the BO algorithm are run to satisfy the convergence criterion \cref{eqn:BO_conv}. Therefore, one optimization at this parameter regime took roughly 87 h of computing time. All optimizations yield different optima: \rev{WF} 1 ($\bv = (-1/6, 0.08,0.08, -1/6, -0.02)$), realizes the greatest power saving ($\Ps=11.8~\%$), while \rev{WF} 2 ($\bv = (0.014,-0.16,-0.14,-1/6,-0.02)$) and \rev{WF} 3 ($\bv=(1/6, -0.07, 0.13, -1/6, 0.06)$) yield power savings of $\Ps=10.8~\%$ and $\Ps=10.4~\%$, respectively. Statistically, using a 95~\% CI and $\zeta^*=2.5~\%$, the difference in power savings is insignificant (the CI around $10.8~\%$ is given by $[6.4~\%, 15.2~\%]$). All power-optimal waveforms also reduce drag substantially: \rev{WF} 1 reduces drag by $\rev{\Dr=}21.1~\%$ when compared to steady driving while \rev{WF} 2 and 3 realize drag reductions of $\rev{\Dr=}20.6~\%$ and $\rev{\Dr=}21.7~\%$.

\rev{
\cite{FoggiRota2023} proposed a different metric than \cref{eqn:DrPs} to evaluate power savings. They accounted for the impossibility of reaching below the laminar state, defining the saving  $P_\mathrm{e}=(1-\mP)/(1-\mP_\ell)$, where $\mP_\ell$ is the power input of the laminar flow state. Using that metric, waveforms 1--3 realize increased savings of $P_\mathrm{e} = 17.6~\%$, $P_\mathrm{e} = 16.1~\%$ and $P_\mathrm{e} = 15.6~\%$, respectively.
}

The finding of multiple minima may have various sources. First, the initial dataset is randomly sampled across the parameter space, which can initially guide the BO loop towards different regions \rev{of the parameter space}. Since the computational resources are limited, the loop cannot gather sufficiently many data points to ensure that the surrogate functions are independent of the initialization. Second, the noisiness can lead to multiple noise-induced minima, especially in flat $\J(\bv)$-landscapes, where noise may be larger than  actual changes in the functional.

\rev{Even if found at different parameters}, all waveforms exhibit similar features. After the minimum flow rate, they follow a steep acceleration phase. As in the power-optimal triangular waveforms, this fast acceleration delays the onset of turbulence \citep{Scarselli2023}. Indeed, the acceleration length of roughly $0.21T$ (measured from the global minimum to the first local maximum in \cref{fig:f_opt}(a)) is similar to the one found in \rev{the} tri-variant approach ($\Ta=0.18T$). Also, as in the triangular waveforms, the maximum velocity is followed by a deceleration phase. Additionally, in contrast to the triangular waveforms, no distinct rest phase exists, and, instead, a second valley-peak combination occurs. 

We analysed the role of the second peak by modifying the best-performing waveform (WF 1 in \cref{fig:f_opt}(a)). Specifically, we replaced the second peak with a rest phase while keeping $\Rem=5160$ (see WF 1a \cref{fig:f_opt}(b)). The power saving for that waveform remains (statistically) identical ($11.2~\%$). A different modification was applied to \rev{WF} 1, namely the minimum Reynolds number was increased from 1500 to 3400 (\rev{WF} 1b in \cref{fig:f_opt}b)). Note that \rev{WF} 1b realized a slightly larger average Reynolds number of $\Rem = 5330$, however, we also normalized the power with the steady power input associated with $\Rem = 5330$. This modification of the waveform results in a substantial decrease of power saving from $11.8~\%$ to $1.9~\%$. The additional power spending in \rev{WF} 1b is linked to the turbulent kinetic energy $\rev{\langle q\rangle_{r, \theta, z}(t)}$, 
shown in \cref{fig:f_opt}(c). As a result of the flatter minimum of \rev{WF} 1b, compared to 1 and 1a, turbulence does not sufficiently decay during the pre-peak phase. Consequently, the onset of turbulence is quicker and peak intensities occur in earlier phases of the deceleration, where the velocity is comparably large. In other words, compared to 1 and 1a, in 1b, a larger amount of fluid has to be displaced at higher turbulence levels, resulting in a larger power consumption during that phase. In \cref{fig:f_opt}(d), we demonstrate that differences in the evolution of the power input of \rev{WF} 1 and 1b mainly occur within the phase of large turbulence intensities.

As all optimal waveforms realize the bounds for at least one coefficient, for \rev{WF} 1, we allowed for larger magnitudes of $1/4$. Using all observations for \rev{WF} 1, the optimization was continued until criterion \eqref{eqn:BO_conv} was fulfilled (31 iterations). At $\bv = (-0.17,0.06,0.1,-0.17,-0.09)$ (found within the interior of the admissible set), the obtained waveform (not shown here) is nearly identical to \rev{WF} 1 and realizes a similar power saving ($\Ps=12.3~\%$).

\rev{Additionally}, we investigated $\Rem = 8600$, drastically increasing the computational effort. Even with our rather coarse computational mesh (see \cref{tab:grid}), single periods roughly took 4 h of computational time, and, additionally, the noise level increases, requiring up to 52 periods to fulfil $\zeta(\overline{\bm{P}}) \leq 2.5~\%$. In total, the 63 iterations of the BO loop needed to satisfy criterion \eqref{eqn:BO_conv} approximately took 65 days of computing time.

\begin{figure}[h!]
    \centering
    \includegraphics[width=1.0\linewidth]{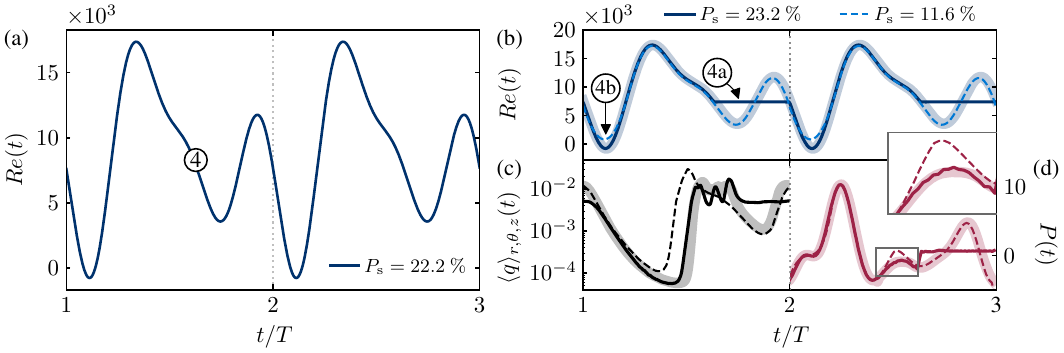 }
    \caption{(a) The power-optimal waveform obtained by the truncated Fourier approach ($\Rem=8600$, \rev{$\Womersley=10$}) where $N=3$ (see \eqref{eqn:fourier}) and $|a_k|, |b_k|\leq 1/4$, $k=1, 2, 3$ (phase-adjusted). (b) \rev{WF} 4 and two modifications thereof (4a and 4b). (c) and (d) evolution of the cross-sectional kinetic energy \rev{(units of $\Um{}^2$)} and the power input (\rev{units of $\mP_\mathrm{b}$;} line styles according to (b)).}
\label{fig:f_opt_8600}
\end{figure}

\Cref{fig:f_opt_8600}(a) shows the power-optimal \rev{WF 4} ($\bv=(0.23,0.25,-0.16,-0.15,-0.05)$) obtained for $N=3$ and $|a_k|,|b_k|\leq 1/4$. Compared to the best-performing waveform at $\Rem=5160$, the power saving doubles ($\Ps =22.2~\%$, \rev{or, if preferred, $P_\mathrm{e}=28.7~\%$}), whereas the drag is reduced by $36.5~\%$. \rev{WF} 4 exhibits similar characteristics to waveforms 1--3, however, the  minimum flow rate becomes negative, corresponding to backflow ($U<0$).

In \cref{fig:f_opt_8600}(b) we show two modifications of \rev{WF} 4, namely 4a and 4b as done earlier for $\Rem=5160$. The findings are shown in \cref{fig:f_opt_8600}(c)--(d) and are analogous: the second peak does not play an important role, the critical aspect of the profile is the deep minimum in the bulk velocity that contributes to sinking turbulence intensities drastically before acceleration.

\begin{figure}[h!]
    \centering
    \includegraphics[width=\linewidth]{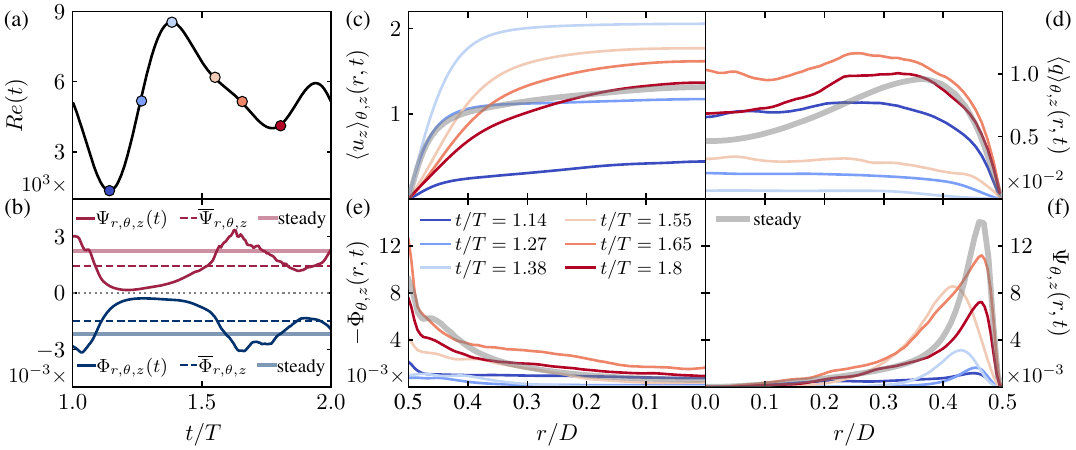}
    \caption{\rev{(a) Optimal \rev{WF} 1 where colours encode the times in ((c)--(f)). (b) Time evolution of production ($\upPsi$) and dissipation ($\upPhi$) in the optimal waveform and steady pipe flow ($\Rem=5160$) in units of $(\Um{}^3/D)$. (c)--(f) Axial and azimuthal averaged axial velocity (units of $\Um$), turbulent kinetic energy (units of $\Um{}^2$), dissipation and production (units of $(\Um{}^3/D)$), averaged over eight periods.}}
    \label{fig:dissprod}
\end{figure}

\begin{figure}[h!]
    \centering
    \includegraphics[width=\linewidth]{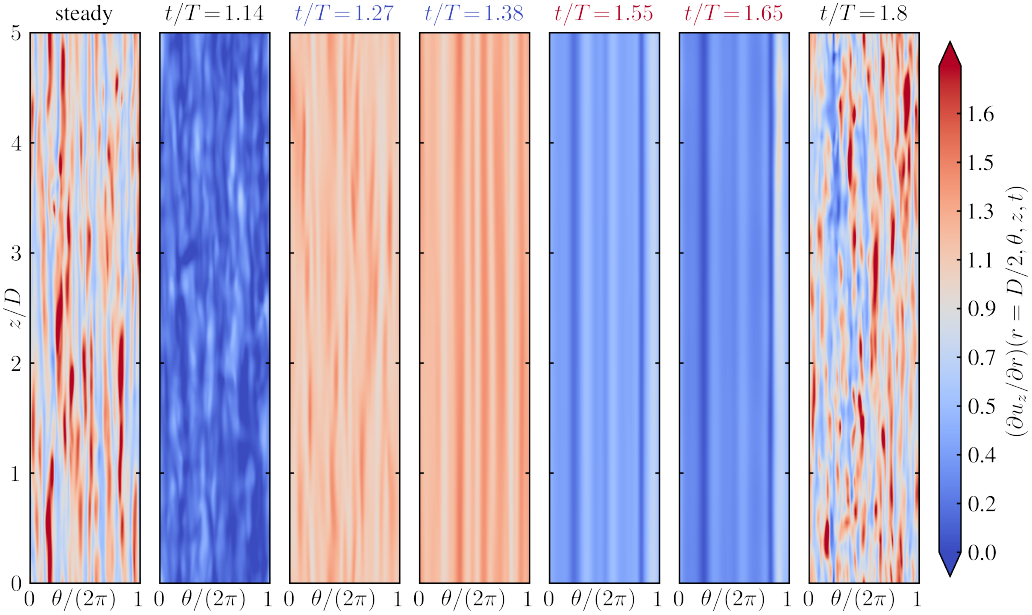}
    \caption{\rev{Instantaneous snapshots of the wall shear stress in units of $\mtw{}_\mathrm{,b}$ in steady pipe flow ($\Rem=5160$, left panel) and in WF 1 for the same phases in the period as in \cref{fig:dissprod}(a).} Blue/red titles are associated with acceleration/deceleration.}
    \label{fig:shear}
\end{figure}

\rev{Lastly, the effect of the Womersley number was investigated. At $\Rem=5160$ and $\Womersley = 10\sqrt2$, we carried out an additional optimization, where $N=3$ and $|a_k|, |b_k|\leq 1/6$, $k=1,2,3$. The optimal WF 5 ($\bv = (1/6, -1/6, 0.03, 0.01,0.10)$) realizes a similar power saving of $10.0~\%$ ($P_\mathrm{e}=14.9~\%$) while reducing drag by roughly $20.6~\%$ (evaluated on our fine grid, see \cref{tab:grid}). The waveform exhibits the same characteristics as WF 1--4, featuring lowest bulk velocities prior to the main acceleration phase. Again, the onset of turbulence is delayed into phases of low Reynolds numbers. In \cref{app:Wo}, we show the optimal waveform for $\Womersley=10\sqrt{2}$
as well as the evolution of the power input and turbulent kinetic energy.
}
\rev{\subsection{Spatio-temporal intra-cycle mechanisms}
We investigated the physical mechanisms that yield energy/drag reductions by computing the production ($\upPsi$) and dissipation ($\upPhi$) of the velocity fluctuations $\bu'$ \citep{feldmann2020}, \par\noindent
\begin{equation}
    \upPsi _{\beta}(t) = -\left(\langle u_r' u_z'\rangle_{\beta} \frac{\partial \langle u_z \rangle_\beta}{\partial r}\right)(t), \quad  \upPhi_\beta(t)= -\left(\frac{1}{\Re} \left\langle \nabla\bm{u}':\nabla\bm{u}'\right\rangle_\beta\right)(t),
\end{equation}
as well as $\langle u_z \rangle_{\theta,z}(r,t)$ and  $\langle q \rangle_{\theta,z}(r,t)$.\\
In figure \ref{fig:dissprod}, we show these quantities for different times within WF 1. Additionally, in \cref{fig:shear} we show instantaneous snapshots of the wall shear stress $({\partial u_z}/{\partial r})(D/2, \theta, z, t)$ in a $\theta$-$z$-plane for a steady case (left panel) and our optimal WF 1 (other panels). 

The minimum of turbulent production occurs right after the minimum bulk velocity (\cref{fig:dissprod}(b)). At this time, turbulent fluctuations are almost completely damped. During the acceleration phase, turbulent production and dissipation are still low ($t/T=1.27$ in \cref{fig:dissprod}(b), (e) and (f), respectively). Therefore, although during the last half of the acceleration, the mean profile has a larger gradient at the wall than the mean profile of the steady case (see \cref{fig:dissprod}(c)), the magnitude of the spatially averaged wall shear stresses (see \cref{fig:shear}) is almost equal between the two. At peak bulk velocity ($t/T=1.38$), the strong acceleration has elongated all velocity streaks close to the wall, resulting in a homogeneous wall shear stress distribution ($t/T=1.38$ in \cref{fig:shear}). During deceleration, turbulent production (and dissipation) increases. Interestingly, first ($t/T=1.55$), the peak of turbulent production is radially localized farther from the wall than the peak production of the steady case (\cref{fig:dissprod}(f)). Thus, turbulent fluctuations are mostly found towards the pipe centre at later pulsation phases ($t/T=1.65$ and $t/T=1.8$ in \cref{fig:dissprod}(d)). At $t/T=1.65$, even though the magnitude of the turbulent fluctuations peaks, the wall shear stress is mostly unaffected (the spatial average amounts to roughly $0.3\mtw{}_\mathrm{,b}$). Only toward the end of the deceleration, the wall shear stress distribution is visually similar to the steady case ($t/T=1.8$). At these phases of the period however, production and (therefore) dissipation decrease, which explains the almost complete dampening of turbulence before the next acceleration phase.

To complement our analysis, in \cref{app:subopt}, we analogously show the quantities as in \cref{fig:dissprod} and \cref{fig:shear}, but for a single harmonic (sine) waveform, with the same the maximum and mean Reynolds number as WF 1. The waveform results in an extra energy consumption of roughly 15~\% and neither drag reduction nor drag increase ($\Dr\approx0$). Due to slower acceleration than in WF 1, we observe smaller wall shear stress in this phase of the period ($t/T=1.4$ and $t/T=1.65$ in \cref{fig:shearw}). The elongated wall shear stress distribution remains, though. During deceleration, both, production and dissipation rise faster (and are substantially larger) than in WF 1 (\cref{fig:dissprodw}(b)). Further, here, the radial localization of the production is similar to the steady case (\cref{fig:dissprodw}(f), $t/T=1.72, 1.82, 1.92$). Therefore, peak turbulent fluctuations appear in earlier phases of deceleration (at larger instantaneous Reynolds numbers), and radial localization resembles the steady case. At late phases of the deceleration ($t/T=1.92$ in \cref{fig:shearw}), the wall shear stress distribution is, like in WF 1,  similar to the steady case, however with a larger magnitude.}

\section{Conclusions}\label{sec:conclusion}

We show that by modulating the bulk velocity in time, while realizing a fixed time-averaged volume flux, \rev{the average wall shear stress} and \rev{net energy} consumption can be greatly reduced when compared to steady driving. This is possible due to the non-linear response of turbulence to unsteady driving: if and to what extent \rev{energy} is saved or drag is reduced is highly dependent on the specific driving waveform \rev{-- motivating optimising for the latter}. 

Solving the optimisation problem using DNS is computationally demanding and results in a noisy estimation of the power input and drag due to the finite-time sampling of the turbulent attractor. Uncertainty in \J can be reduced by running longer simulations (i.e., averaging over more periods) at the cost of increasing computing times. In gradient-based methods, including adjoint optimization, noise must be small in each iteration to enable reasonable gradient approximations. BO deals by design with noisy objective functionals by modelling noise with Gaussian processes. Hence, single functional evaluations can be noisy and thus cheap. As more observations are added, the surrogate model
is iteratively refined and the uncertainty in the surrogate is much smaller than the noise in the functional. We note that BO had already been used in fluid problems, e.g., for flow control by \cite{Morita2022,Mallor2023}, who showed that a few function evaluations are needed to converge to global optima. Our work unveils the ability of BO to deal with (large) noise levels and highlights its prowess to optimize average turbulent flow quantities. Generally, we  expect BO to significantly outperform gradient-based algorithms in the optimisation of ergodic properties of chaotic systems such as turbulence.  

We find that, to save energy, it is mandatory to decrease the turbulence intensity prior to the acceleration phase, thereby delaying the onset of turbulence, in agreement with \citet{Scarselli2023}. Independently of the parametrization method considered (uni/tri-variant or Fourier-series), all found optimal waveforms have a steep acceleration phase in common. As noted by \citet{Moron2022} in the context of transitional pulsatile pipe flow, and \citet{Scarselli2023} in the case of fully turbulent flow, at these Womersley numbers, a fast acceleration is related to a rapid turbulence collapse. This is ultimately rooted in the time delay between the driving pressure gradient and the turbulence response, that for these flow parameters was shown to be roughly $T/4$ \citep{Moron2024b}. After modifying our optimal waveforms with different techniques, we show that changing the acceleration phase slope \rev{(and thereby not delaying the onset of turbulence)} has the largest detrimental impact on \rev{energy consumptions}. \rev{Compared to a single harmonic waveform, the optimal waveform exhibits a substantially larger time shift between the peak bulk velocity and the peaks of dissipation and production. Further, in optimal waveforms, both, turbulent kinetic energy and production are more radially localized toward the pipe centre, resulting in lower spatially and temporally averaged wall shear stresses.}

In conclusion, thanks to the use of Bayesian optimisation, coarse meshes and a state-of-the-art GPU DNS code, we solve the optimisation problem in a broad parametric regime and computationally efficiently. We consider a large space of continuous waveforms by optimising a small number of Fourier coefficients of the driving waveform (see \cref{eqn:fourier}). Here, at $\Rem=8600$, the optimal waveform saves up to $23~\%$ of energy and reduces drag by $37~\%$, clearly outperforming triangular waveforms at the same Reynolds number \citep[][reported a power saving of $9~\%$ and a drag reduction of $27~\%$ at this $\Rem$]{Scarselli2023}. We \rev{only considered two ($\Womersley\in\{10, 10\sqrt{2}\}$)} and investigated three average Reynolds numbers ($\Rem\in\{4300, 5160, 8600\}$). Hence, a large parameter space remains to be explored in the future. In addition, for computational reasons, we restricted the number of Fourier coefficients and their magnitude, imposing a strong locality to the optima. However the, presented savings are already higher than many flow control techniques documented in the literature \citep[e.g.][]{FoggiRota2023, FoggiRota2023b, Scarselli2023}. Finally, we note that in practical applications, further steps are needed to realize the power savings. For example, power gains occur in part of the period where the fluid is decelerated and storing this energy will unavoidably result in losses.

\newpage

\begin{bmhead}[Acknowledgements]
We appreciate fruitful discussions with Dr.~Jos\'e M. Lopez (University of M\'alaga) and the members of the fluid modeling and simulation group at ZARM.
\end{bmhead}

\begin{bmhead}[Funding]
Support from the Deutsche Forschungsgemeinschaft (DFG, German Science Foundation) under grant number 540652448 is gratefully acknowledged.
\end{bmhead}

\begin{bmhead}[Declaration of interests]
The authors report no conflict of interest.
\end{bmhead}

\begin{bmhead}[Data availability statement]
The data that support the findings of this study will be made openly available in \pangaea.
\end{bmhead}

\begin{bmhead}[Author ORCIDs]
Felix~Kranz, \url{https://orcid.org/0009-0006-0974-450X}\\
Daniel~Morón, \url{https://orcid.org/0000-0001-7057-0082}\\
Marc~Avila, \url{https://orcid.org/0000-0001-5988-1090}
\end{bmhead}

\bibliographystyle{jfm}
\bibliography{bib}

\newpage
\appendix
\section{Gradient-based optimisation}\label{app:gds}
In this appendix we describe the results of using a gradient-based optimisation method. In particular we consider an ordinary gradient-based sequential least-squares program (SLSQP, \cite{Schittkowski1982, kraft1988}), and study its performance with respect to the desired level of standard error $\zeta^{*}$. We implement the version from \href{https://github.com/scipy/scipy}{\texttt{scipy}}'s optimise package.\\
The SLSQP method is a quasi-Newton method, suited for non-linear constrained optimisation problems. It follows the basic update strategy $\bv{}^{(j+1)} = \bv{}^{(j)} + \eta{}^{(j)}\bs{}^{(j)}$ where $\eta{}^{(j)}$ and $\bs{}^{(j)}$ denote the step size and search direction at iterate $j$. In each iteration, the search direction is computed by solving the least squares sub-problems \citep{Schittkowski1982} arising from a factorization of a Broyden-Fletcher-Goldfarb-Shanno-approximation to the Hessian  
(BFGS, \cite{broyden1970, Fletcher1970, Goldfarb1970, Shanno1970}).

\begin{figure}[h!]
    \centering
    \includegraphics[width=1.0\linewidth]{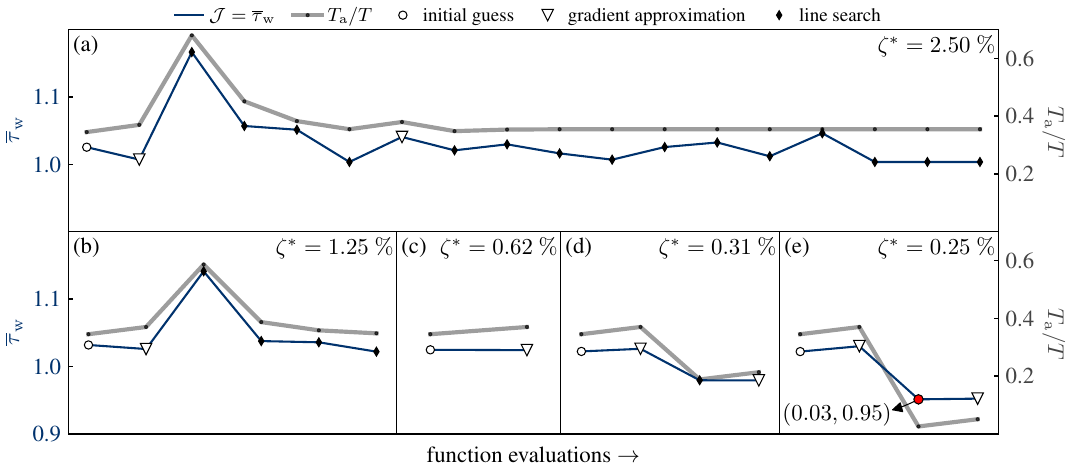}
    \caption{The evolution of the optimisation process for different choices of the admissible standard error $\zeta^* \in \{2.5, 1.25, 0.625, 0.3125, 0.25\}~\%$. Blue lines indicate the \rev{wall shear stress} obtained when the acceleration time is chosen according to the grey lines (uses right hand side labels), where gradient-approximations are indicated by downward triangles and line searches are indicated by diamonds.}
\label{fig:slsqp}
\end{figure}

The Armijio condition \citep{Armijo1966} is used in the line search to guarantee an appropriate step size $\eta{}^{(j)}$. A detailed description of the SLSQP method can be found in \cite{Schittkowski1982}, \cite{kraft1988} or \cite{Ma2024} and the references therein.\\
We assessed the performance in terms of feasibility, computational effort and robustness of the SLSQP method using the restricted (uni-variant) triangular waveform described in \cref{sec:waveforms}.  Here, choosing $\Rem=4300$, $\Re^+=9400$ and $\Re^-=1600$ (see \cref{fig:fig1}(a)), we aim to identify the $\mtw$-optimal acceleration time. The acceleration time is bounded by $0.01T$ and $0.68T$ ($\Ta \in \mathcal{Q}=[0.01, 0.68]T$) in order to obtain continuous waveforms and to ensure that $\Rem$ is realized. The gradient ${\partial \mathcal{J}}/{\partial \Ta}$ (used for the computation of the search direction) is approximated using first-order finite differences with a step size of $2.5\cdot10{}^{-2}T$. The optimisation is started with $\bv{}^{(0)}=0.345$ and the optimisation loop is terminated if either $|\mathcal{J}(\bv{}^{(j)})-\mathcal{J}(\bv{}^{(j-1)})|\leq \epsilon $ or $\|\bs{}^{(j)}\|_2\leq\epsilon$, where the tolerance is chosen as $\epsilon=5\cdot10{}^{-3}$.\\
For different values of $\zeta^*$ \rev{and initially $n=3$}, the evolution of the function value $\mtw$ (blue lines) and the corresponding acceleration time (grey lines) as the SLSQP-optimisation loop progresses are shown in \cref{fig:slsqp}{(a)--(e)}. For $\zeta^*=2.5~\%$ (\cref{fig:slsqp}\textbf{(a)}), the gradient at the initial guess points towards larger acceleration times ($\Ta$ is increased and $\mtw$ decreases). In fact, the gradient is so large that the first line search iteration moves towards the upper bound of $\mathcal{Q}~(\Ta=0.68T)$, increasing the function value to $\mtw = 1.17$. Subsequent line search iterations find an acceleration time where, compared to the initial value of $\mtw=1.025$, the mean \rev{wall shear stress} is decreased to $\mtw = 1.004$. Even though being close to the initial guess, at $\Ta=0.355$ the now gradient points towards shorter accelerations (second $\nabla$-marker in \cref{fig:slsqp}\textbf{(a)}), which would indicate a minimum between $0.345$ and $0.355$. Subsequently, the line search fails to find a step that decreases the function value for 11 consecutive function calls, and the optimisation loop is terminated. These line searches unveil that the underlying optimisation problem exhibits features of ill-posedness: minor changes in the acceleration time yield large changes in the function value. Thus, gradient approximation and line searches are prone to following those noise-induced changes rather than capturing the overall trend of the function. This motivates
the question of how many periods to run in order to capture the underlying function trend.\\
For $\zeta^*=1.25~\%$, the gradient at $\bv^{(0)}$ also points towards larger accelerations (if not so strongly) and the line search finds such a small step size that $\J(\bv^{(0)})$ and $\J(\bv^{(5)})$ almost coincide and the loop terminates. Reducing $\zeta^*$ to $0.625~\%$ in \cref{fig:slsqp}(c), leads to an almost zero gradient at the initial guess and the loop is terminated by the $\|\mathbf{s}^{(j)}\|_2\leq \epsilon$ criterion. For $\zeta^*=0.3125~\%$ (\cref{fig:slsqp}(d)), the gradient points towards shorter acceleration times and the line search finds a suitable step size after only one iteration, indicating the gradient is now capturing the trend of the function. Lastly, for $\zeta^*=0.25~\%$ the slope towards smaller acceleration times is even larger, so that first line search iteration (red marker in \cref{fig:slsqp}\textbf{(e)}) finds $\J(U_{0.025}(t))=0.951$.\\
Even at $\zeta^* = 0.25~\%$, which in our case is equivalent to running up to 98 periods, the SLSQP method evidently fails to find minima: at $\Ta=0.01T$, using $\zeta^*=0.25~\%$, a mean \rev{wall shear stress} of $\mtw=0.945$ is obtained, which is statistically significantly smaller than the obtained minimum of $\mtw=0.951$ (the 95~\% confidence interval (CI) of 0.951 is given as $[0.946, 0.956]$). Clearly, further reducing $\zeta^*$ is not a viable strategy to find ``true" minima, as it comes with extensive computational effort. In fact, the four function evaluations from \cref{fig:slsqp}(e) took 4.5 days. Assuming that the optimisation would converge for a smaller standard error of, for example, $0.125~\%$, according to \cref{eqn:ste} four times as many periods (roughly 400) are necessary, which would lead to a computational time of 4.5 days per function evaluation. We conclude that, for the multidimensional optimisation problems proposed here, where the number of function evaluations in general scales with $d^2$, gradient-based methods like the SLSQP method are computationally infeasible.\\
Increasing the step size for the finite difference gradient approximation can be a way to overcome noise-induced local minima. However, the chosen value of $2.5\cdot\num{e-2}$ is already relatively large and increasing it further reduces the accuracy substantially (i.e. we would be unable to capture short term trends in \Js).

\newpage 
\rev{\section{Effect of the Womersley number}\label{app:Wo}
For $\Rem=5160$ and $\Womersley=10\sqrt{2}$, we show the optimal waveform in \cref{fig:Wo14}(a). Here, a power saving of 10.0 \% is realized while drag is reduced by 20.7 \%. The waveform shows the same characteristics as WF 1--4. The pre-peak velocity is reduced substantially, so that the onset of turbulence is delayed (see \cref{fig:Wo14}(b)). As before, quick acceleration needs large instantaneous pressure gradients (\cref{fig:Wo14}(c)) and therefore the instantaneous power input is up to 14 times larger than in steady conditions (\cref{fig:Wo14}(e)).
\begin{figure}[h!]
    \centering
    \includegraphics[width=\linewidth]{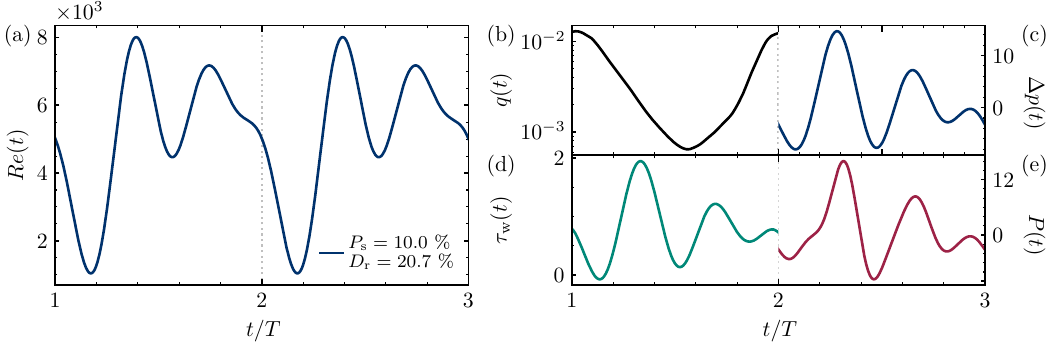}
    \caption{ \rev{(a) The power-optimal waveform obtained by the truncated Fourier approach ($\Rem=5160$, $\Womersley=10\sqrt2$) where $N=3$ (see \eqref{eqn:fourier}) and $|a_k|, |b_k|\leq 1/6$, $k=1, 2, 3$ (phase-adjusted). (b) The evolution of the cross-sectional kinetic energy (units of $\Um{}^2$) and the power input (units of $\mP_\mathrm{b}$). (c)--(e) shows the evolution of the pressure drop, the wall shear stress and the power input (in units of the steady state counterparts), respectively.}}
    \label{fig:Wo14}
\end{figure}
}

\rev{\section{Robustness of optimal waveforms}\label{app:robust}
First, we analyse the sensitivity with respect to the spatial and temporal resolution. As shown in table \ref{tab:grid}, for waveform 1--3 ($\Rem=5160$), we conducted additional simulations with significantly finer resolutions while reducing the time step roughly tenfold (see \cref{tab:grid}). We note that our code does not adjust the time step based on the Courant-Friedrichs-Lewy (CFL) number, but based on an error threshold of the predictor-corrector loop for the non-linear term. The observed changes in power savings are insignificant for any waveform when considering the 95 \% CI with the acceptable standard error of 2.5 \%. In WF 1, power saving is increased to 14.5 \% (originally 11.8 \%), while in WF 2 the power saving is nearly identical (10.1 \% vs. 10.8 \% originally) and in WF 3, the saving is slighly reduced to 8.5 \% (10.4 \% originally). In WF 4, power savings are slightly increased to 23.4~\% (originally 22.2 \%).  We stress that satisfying such strict spatial and temporal resolutions during the whole optimization process is infeasible: for the $\Rem=5160$ case the single function evaluations take roughly 2.5 days, while they take 9 days for the $\Rem=8600$ case.\\
Second, we examine the effect of the body force \eqref{eqn:forcing}. On our finest grid (see \cref{tab:grid}), we have conducted additional simulations of the optimal waveforms 1--3 without any forcing term ($A_\mathrm{f}=0$). In figure \ref{fig:forcing}(a), we show the evolution of the turbulent kinetic energy during the first 3 periods for waveforms 1--3. While in waveforms 1 and 2 turbulence, turbulence is sustained, WF 3 shows exponential decay of turbulence. Thus, in WF 3, laminar conditions are approached, resulting in an unrealistic power saving of roughly 40 \%. Drag is reduced by 44 \%. In the optimization loop, waveforms that relaminarise would always be preferable over those that do not -- even if in reality, they might perform worse. In the waveforms 1 and 2, the forcing has minimal impact on power savings: figure \ref{fig:forcing}(b) shows the time integrated difference between the power input $P(t)$ with and without forcing, $\delta P(t)$. In the majority of the period, power inputs with and without forcing are almost equal ($\delta P(t)\approx 0$). Towards the end of the period, they slightly diverge, however $\delta P(T)$ (which is the difference $\Ps$ with and w/o forcing), is insignificant ($\approx 4~\%$). In WF4 ($\Rem=8600$), turbulence is sustained. The change in the power saving is insignificant ($\Ps = 22.1~\%$ vs. $\Ps = 22.2~\%$ originally).\\
\begin{figure}
    \centering
    \includegraphics[width=\linewidth]{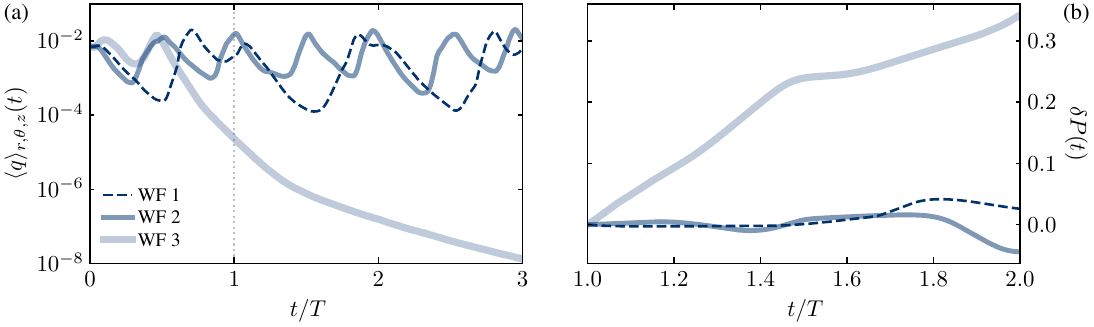}
    \caption{\rev{(a) The evolution of the turbulent kinetic energy (in units of $\Um{}^2$) for waveforms 1--3 ($\Rem=5160$) without forcing ($\Af=0$). (b) The time integrated difference of the power inputs $P(t)$, $\delta P(t)$, with ($\Af=1.15\cdot10{}^{-3}$) and w/o forcing.}}
    \label{fig:forcing}
\end{figure}
The optimal waveforms are robust with respect to the pipe length. For waveforms 1--3, we conducted simulations on a longer pipe $(L=10D)$. Power savings fall in line with the ones obtained in the short pipe ($\Ps=12.6~\%$ for waveform 1, $\Ps=10.9~\%$ for \rev{WF} 2 and $\Ps=10~\%$ for \rev{WF} 3). \\
Lastly, we have also confirmed the robustness with respect to different initial conditions. For waveforms 1--3, instead of using an already turbulent initial condition, obtained from a steady run, we initialize the flow laminar ($u_z=1-r^2$) and introduce a (strong) localized perturbation of the form
\begin{align*}
    u_r &= 0.2(1-r^2)^2\exp\left(-10\sin^2(\pi z/L)\right)\sin(\theta)\\
    u_\theta &= 0.2\left((1-r^2)^2-4r^2(1-r^2)\right)\exp\left(-10\sin^2(\pi z/L)\right)\cos(\theta).
\end{align*}
This unveils that the waveforms are memoryless, in the sense that even strong initial conditions take no influence on the flow behaviour after the first period. Precisely, for waveforms 1, 2 and 3 (see \cref{fig:f_opt}), we now obtain power savings of 13 \%, 9.6 \% and 9.2 \%, respectively. These differences are insignificant considering the standard error of 2.5 \% that was used during optimization. \\
}
\newpage
\rev{\section{Spatio-temporal intra-cycle mechanics in sub-optimal waveforms}\label{app:subopt}
Here, we show the figures for the intra-cycle mechanism of a single harmonic waveform, namely \cref{fig:dissprodw} and \cref{fig:shearw}. 

\begin{figure}[h!]
    \centering
    \includegraphics[width=\linewidth]{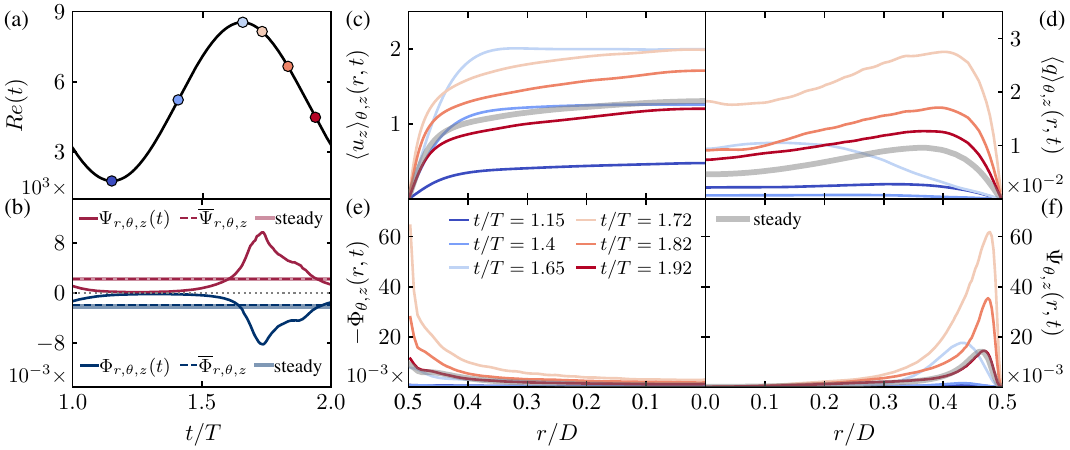}
    \caption{\rev{(a) Single harmonic waveform ($\Rem=5160$, $\Re^+=8550$, $\Re^- = 1779$) where colours encode the times in ((c)--(f)). (b) Time evolution of production ($\upPsi$) and dissipation ($\upPhi$) in this waveform and steady pipe flow ($\Rem=5160$) in units of $(\Um{}^3/D)$. (c)--(f) Axial and azimuthal averaged axial velocity (units of $\Um$), turbulent kinetic energy (units of $\Um{}^2$), dissipation and production (units of $(\Um{}^3/D)$).}}
    \label{fig:dissprodw}
\end{figure}

\begin{figure}[h!]
    \centering
    \includegraphics[width=\linewidth]{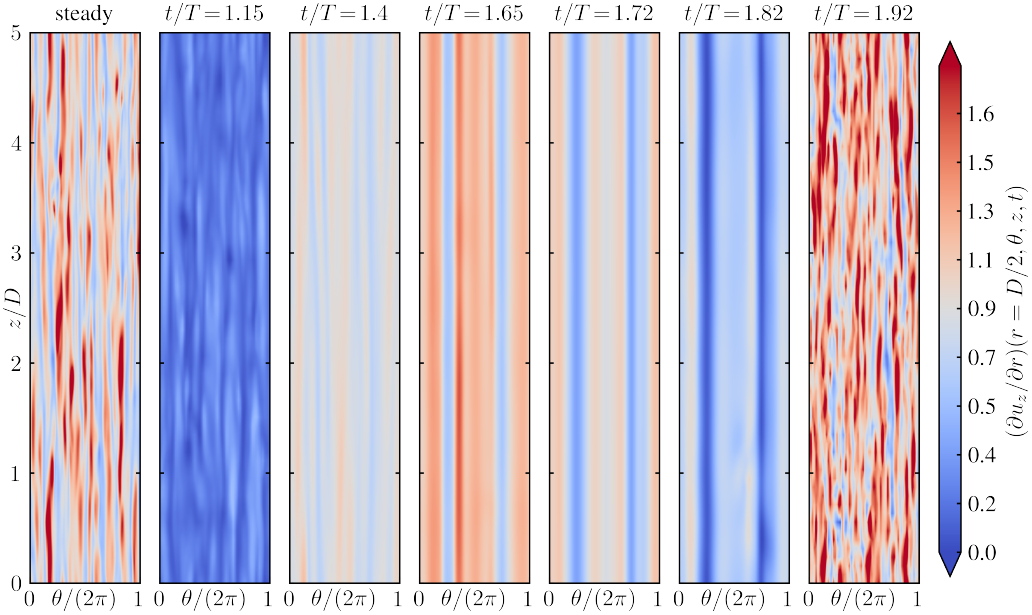}
    \caption{\rev{Instantaneous snapshots of the wall shear stress in units of $\mtw{}_\mathrm{,b}$ in steady pipe flow ($\Rem=5160$, left panel) and in the single harmonic waveform for the same phases in the period as in \cref{fig:dissprodw}(a).}}
    \label{fig:shearw}
\end{figure}

}

\end{document}